\documentclass[11pt,letterpaper]{article}
\usepackage[T1]{fontenc}
\usepackage[utf8]{inputenc}
\usepackage{lmodern}
\usepackage[margin=1in]{geometry}
\usepackage{amsmath,amssymb,amsthm}
\usepackage[colorlinks=true,linkcolor=blue,citecolor=blue,urlcolor=blue]{hyperref}
\hypersetup{pdftitle={Computing EF1 Allocations under Budget Constraints in Polynomial Time},
  pdfauthor={Tingwei Hu, Lili Mei, Guochuan Zhang}}
\usepackage{enumitem}
\usepackage{float}
\floatstyle{ruled}
\newfloat{algorithm}{htbp}{loa}
\floatname{algorithm}{Algorithm}
\usepackage{needspace}
\newtheorem{theorem}{Theorem}[section]
\newtheorem{lemma}[theorem]{Lemma}
\theoremstyle{definition}
\newtheorem{definition}[theorem]{Definition}
\newcommand{\OPT}{\operatorname{OPT}}
\newcommand{\EFValid}{\ifmmode\text{\normalfont\textsc{EF1-Valid}}\else{\normalfont\textsc{EF1-Valid}}\fi}
\newcommand{\EFInvalid}{\ifmmode\text{\normalfont\textsc{EF1-Invalid}}\else{\normalfont\textsc{EF1-Invalid}}\fi}
\newcommand{\SSum}{\ifmmode\text{\normalfont\textsc{Subset Sum}}\else{\normalfont\textsc{Subset Sum}}\fi}
\DeclareMathOperator*{\argmin}{arg\,min}
\title{Computing EF1 Allocations under Budget Constraints in\\ Polynomial Time}
\author{Tingwei Hu$^1$, Lili Mei$^2$, Guochuan Zhang$^1$\\[1ex]
\small $^1$College of Computer Science, Zhejiang University, Hangzhou, China\\
\small $^2$School of Cyberspace, Hangzhou Dianzi University, Hangzhou, China\\[1ex]
\small\href{mailto:tingweihu@zju.edu.cn}{\texttt{tingweihu@zju.edu.cn}}\quad
\href{mailto:meilili@hdu.edu.cn}{\texttt{meilili@hdu.edu.cn}}\quad
\href{mailto:zgc@zju.edu.cn}{\texttt{zgc@zju.edu.cn}}}
\date{}
\begin{document}
\maketitle
\begin{abstract}
We study the allocation of indivisible goods to agents with budgets and additive valuations. Each agent compares its bundle with every budget-feasible subset of another agent's bundle or of the unallocated goods. We give a strongly polynomial algorithm for computing an allocation that is envy-free up to one good (EF1), even when agents have different valuations and budgets. Deciding whether a given allocation satisfies EF1 is coNP-complete. Equivalently, deciding whether it violates EF1 is NP-complete. When all agents have the same valuation, we can compute an EF1 allocation with welfare at least $(1/2-\varepsilon)$ times the maximum feasible welfare in time polynomial in the input length and $1/\varepsilon$, for any $0<\varepsilon<1/2$.

Finding an allocation that is envy-free up to any good (EFX) is NP-hard even for one agent. We give polynomial-time EFX algorithms when each agent values goods in proportion to their sizes or when the number of distinct positive item values per agent is bounded by a constant.
\end{abstract}

\section{Introduction}\label{sec:s1}

We study the allocation of indivisible goods to agents with individual budgets and additive valuations. A good has a common size but may have different values to different agents. Each agent must receive a bundle whose total size fits its budget. The unallocated goods form a bundle called the \emph{charity}.

In classical fair division, an allocation is envy-free up to one good (EF1) if removing some good from an envied bundle eliminates the envy~\cite{r3,r4}. Under budget constraints, each agent makes this comparison with every nonempty subset of another bundle that fits its own budget~\cite{r14}. We impose the same requirement on the charity, since otherwise leaving all goods unallocated would trivially satisfy EF1 among agents. Envy-freeness up to any good (EFX) requires that removing \emph{any} good from the subset eliminate the envy.

The first guarantees for this model were approximate. Wu, Li, and Gan~\cite{r14} showed that maximizing Nash social welfare gives $1/4$-EF1. For identical additive valuations and possibly different budgets, Gan, Li, and Wu~\cite{r15} gave a polynomial-time $1/2$-EF1 algorithm, and Barman et al.~\cite{r2} obtained EF2 in polynomial time, allowing two goods to be removed in each comparison. Barman et al.~\cite{r5} subsequently gave a pseudo-polynomial algorithm for computing an allocation that satisfies a stronger fairness condition, feasible EFx (FEFx). For identical valuations, Hu et al.~\cite{r6} also showed that some EF1 allocation achieves at least half of the maximum feasible welfare. These results left open whether an exact EF1 allocation could be computed in polynomial time, even for identical valuations and unequal budgets~\cite{r1,r2}.

\subsection{Contributions}\label{sec:s1-1}

\begin{itemize}[leftmargin=*,itemsep=3pt,topsep=3pt]
\item \textbf{Computing EF1 (Theorems~\ref{theorem:s3-2} and~\ref{theorem:s4-1}).} A strongly polynomial algorithm computes an EF1 allocation for arbitrary budgets and additive valuations. For $n$ agents and $m\ge1$ goods, it can be implemented in $\widetilde O(n^2m^4)$ arithmetic operations.
\item \textbf{Verifying EF1 (Theorems~\ref{theorem:s5-2} and~\ref{theorem:s5-3}).} Deciding whether a given feasible allocation is EF1 is coNP-complete, and deciding whether it violates EF1 is NP-complete. Both results hold even for one agent with charity, and also for two agents with identical valuations and empty charity.
\item \textbf{Welfare under EF1 (Theorem~\ref{thm:welfare-near-half}).} With identical valuations, an EF1 allocation attaining at least a $(1/2-\varepsilon)$ fraction of the maximum feasible welfare can be computed in time polynomial in the input length and $1/\varepsilon$, for every rational $0<\varepsilon<1/2$.
\item \textbf{Computing EFX (Theorems~\ref{thm:efx-search}, \ref{theorem:s6-2}, and~\ref{theorem:s6-4}).} Finding an EFX allocation is NP-hard even for one agent with charity. If each agent values goods in proportion to their sizes, an EFX allocation can be found in $O((n+m)^2)$ operations. If each agent has at most a fixed number of distinct positive item values, a polynomial-time algorithm computes an allocation satisfying the stronger FEFx condition.
\end{itemize}

\subsection{Technical overview}\label{sec:intro-technical}

\paragraph{Constructing EF1 with approximate knapsack solutions.}
EF1 requires a comparison with every subset of a bundle that fits the observer's budget. Our algorithm builds bundles from the charity and maintains EF1 as it goes. The link between approximation and exact fairness is that removing a most valuable good from a set of at most $m$ goods leaves at most a $(1-1/m)$ fraction of its value. Thus, if an agent's current utility is at least this fraction of the best value it could obtain from a candidate bundle, EF1 holds for every feasible subset of the candidate. Lemma~\ref{lem:ef1-sufficient} shows how an approximate knapsack solution certifies this inequality.

The algorithm first gives a feasible, positively valued singleton to an agent of zero utility whenever possible. It then selects a candidate from the charity and examines it once for each agent with positive utility. If the candidate exceeds an agent's budget, a knapsack call either certifies EF1 or replaces the candidate by a feasible subset. Once the candidate fits, the algorithm deletes goods of least value to that agent until removing a most valuable good would eliminate any envy. Each later agent sees a subset of the previous candidate, so earlier comparisons remain valid. The last agent to select or shrink the candidate receives it and strictly improves its utility. When no candidate is selected, the knapsack bounds certify EF1 towards the charity as well.

\paragraph{Bounding the number of updates.}
A replacement that ends with a deletion can increase the recipient's utility by an arbitrarily small amount. We show that each such increase crosses a sum of consecutive entries in that agent's sorted list of item values. There are only $O(m^2)$ such sums and utilities never decrease. Every other replacement increases utility by a factor greater than $1/(1-1/(2m))$. To count these increases independently of the magnitudes of the values, we cover the attainable positive utilities by a union of at most $m$ intervals whose length on a logarithmic scale is at most $(m-1)\ln 2$. Together the two counts give fewer than $2m^2$ bundle changes per agent. Standard knapsack profit scaling suffices for strong polynomiality. Appendix~\ref{app:oracle} gives the faster implementation used in Theorem~\ref{theorem:s4-1}.

\paragraph{Welfare under identical valuations.}
With identical valuations, the welfare proof splits an optimal allocation according to which goods our algorithm assigned and which it left in charity. The assigned goods have the same value whoever holds them. For the remaining goods, the stopping condition bounds what each agent could gain from the charity in terms of its own utility. Adjusting the accuracy of the knapsack calls makes the resulting welfare guarantee arbitrarily close to one half.

\paragraph{Finding EFX allocations.} The hardness reduction encodes the answer to a \SSum{} instance in a single designated good. Whether the agent receives that good reveals the answer for every EFX output. The two tractable cases use different properties of the valuations. Under proportional valuations, we choose an agent of minimum current load and assign it a good of maximum size among those that fit. The last good it receives is then among its smallest. Removing a smallest good leaves the largest possible value, so this is the critical comparison for EFX. With a fixed number of positive value classes, enumerating the item counts in each class solves the knapsack subproblem exactly and bounds the number of distinct utilities an agent can attain.

\subsection{Related work}\label{sec:s1-2}

\paragraph{EF1 and EFX without budget constraints.} EF1 was formalized by Budish \cite{r3}. The earlier algorithm of Lipton et al. \cite{r4}, based on eliminating envy cycles, already yields an EF1 allocation in polynomial time for additive valuations. Caragiannis et al. \cite{r7} show that maximizing Nash social welfare gives both EF1 and Pareto optimality, and introduce the stronger notion of EFX. These results concern the classical setting, where every bundle is feasible for every agent.

Complete EFX allocations exist in several cases. Plaut and Roughgarden \cite{r8} study EFX under general valuations, including the cases of identical valuations and two agents. Chaudhury, Garg, and Mehlhorn \cite{r9} prove existence for three agents with additive valuations.

\paragraph{Charity and partial EFX allocations.} Allowing unallocated items can help achieve EFX. Chaudhury et al. \cite{r10} show that an EFX allocation exists with at most $n-1$ unallocated items, with no agent preferring the unallocated set to its own bundle. Berger et al. \cite{r11} obtain an EFX allocation for four agents with at most one unallocated item, again without envy towards the unallocated set.

Other work studies partial EFX allocations with high Nash social welfare. Caragiannis, Gravin, and Huang \cite{r12} show that leaving items unallocated can achieve EFX with at least half the maximum Nash social welfare. Feldman, Mauras, and Ponitka \cite{r13} characterize optimal tradeoffs between approximate EFX and Nash social welfare. Nash social welfare is the geometric mean of the agents' values.

\paragraph{Budget and generalized assignment constraints.} Beyond the approximation guarantees discussed above, Wu, Li, and Gan~\cite{r16} give exact polynomial-time EF1 algorithms for two agents and for agents with identical budgets and valuations. For identical valuations, the EF2 algorithm of Barman et al.~\cite{r2} guarantees EF1 when all goods have the same size, value, or value-to-size ratio. Barman et al.~\cite{r5} allow sizes as well as valuations to vary across agents. For their stronger FEFx condition, they give a fully polynomial-time approximation scheme and prove that exact computation is NP-hard. Section~\ref{sec:s6-1} explains the distinction between FEFx and the EFX condition studied here. Elkind, Igarashi, and Teh~\cite{r1} study budget constraints for chores, and Cohen et al.~\cite{r17} consider goods that must be assigned irrevocably as they arrive.

\paragraph{Other feasibility constraints.} Suksompong \cite{r18} surveys constrained fair division, including budget constraints. For example, Biswas and Barman \cite{r19} give polynomial-time EF1 algorithms under category-wise cardinality constraints with additive valuations. Such constraints limit the number of items from each category, whereas budgets limit a weighted sum of item sizes and may differ across agents.

\paragraph{The price of fairness.} The price of fairness measures the loss in optimal welfare caused by a fairness requirement. Early studies by Bertsimas, Farias, and Trichakis \cite{r20} and by Caragiannis et al. \cite{r21} examined this welfare loss. Bei et al. \cite{r22} study the price of fairness for indivisible goods under several fairness notions. For normalized additive valuations, their lower bound and the upper bound of Barman, Bhaskar, and Shah \cite{r23} give a tight $\Theta(\sqrt{n})$ bound for EF1. In the two-agent normalized case, Li et al. \cite{r24} establish the exact EF1 price of $8/7$. Under normalization, all agents assign the same value to the set of all goods. This assumption matters for welfare comparisons because rescaling an individual agent's values preserves envy comparisons but can change the allocation maximizing total welfare.

Hu et al. \cite{r6} extend the price-of-fairness analysis to budget constraints. For homogeneous valuations, they prove an upper bound of 2 and lower bounds approaching $2-1/n$, for both EF1 and EFX. Their upper bounds prove that high-welfare allocations exist but do not give a polynomial-time algorithm.

\paragraph{Optimizing welfare subject to fairness.} Computing an arbitrary fair allocation and finding the best fair allocation are different computational tasks. Aziz et al. \cite{r25} study welfare maximization under EF1, including hardness even for two agents and pseudo-polynomial algorithms for a fixed number of agents. Bu et al. \cite{r26} study welfare approximation under EF1 and EFX for both normalized and unnormalized valuations. These works concern optimization over fair allocations in the classical setting.

\subsection{Organization of the paper}\label{sec:s1-3}

Section~\ref{sec:s2} defines the model and fairness requirement. Section~\ref{sec:s3} presents the EF1 algorithm and proves its correctness. Section~\ref{sec:s4} establishes its polynomial running time. Section~\ref{sec:s5} studies EF1 verification complexity. Section~\ref{sec:s6} proves NP-hardness of finding an exact EFX allocation and gives polynomial-time algorithms in two special cases. Section~\ref{sec:s7} applies the EF1 algorithm to social welfare under homogeneous valuations.

\section{Preliminaries}\label{sec:s2}

Let $N=\{1,\ldots,n\}$ be the set of agents and let $M$ be a set of $m$ indivisible items, where $n\ge1$.
Each item $g\in M$ has size $s(g)\in\mathbb{Q}_{>0}$. Each agent $i\in N$ has a budget $B_i\in\mathbb{Q}_{\ge0}$ and a value $v_i(g)\in\mathbb{Q}_{\ge0}$ for each item. Valuations are additive. For every $T\subseteq M$, write
\[
s(T)=\sum_{g\in T}s(g),\qquad v_i(T)=\sum_{g\in T}v_i(g).
\]
The valuations are \emph{homogeneous} if $v_i(g)=v_j(g)$ for all agents $i,j$ and items $g$. Otherwise, they are \emph{heterogeneous}. Item sizes are common to all agents throughout the paper.

An allocation profile is $A=(X_1,\ldots,X_n)$, where the bundles are pairwise disjoint subsets of $M$.
The charity is $C=M\setminus\bigcup_{i\in N}X_i$. An allocation is \emph{budget-feasible} if $s(X_i)\le B_i$ for every agent $i$.
There is no size constraint on the charity.

\begin{definition}[EF1 under budget constraints]\label{definition:s2-1}
Consider a budget-feasible allocation. Agent $i$ is EF1 towards agent $j$, where $j\ne i$, if, for every nonempty subset $T\subseteq X_j$ with $s(T)\le B_i$, there exists an item $g\in T$ such that
\[
v_i(X_i)\ge v_i(T\setminus\{g\}).
\]
Agent $i$ is EF1 towards the charity if, for every nonempty subset $T\subseteq C$ with $s(T)\le B_i$, there exists an item $g\in T$ satisfying the same inequality. The allocation is EF1 if every agent is EF1 towards every other agent and every agent is EF1 towards the charity.
\end{definition}

Since valuations are additive, each of these EF1 conditions is equivalent to
\[
v_i(X_i)\ge v_i(T)-\max_{g\in T}v_i(g)
\]
for every such nonempty subset $T$. The right-hand side is the value remaining after removing an item of maximum value to agent $i$. Empty comparison sets impose no fairness condition. We say that agent $i$ \emph{envies} a set $T$ if $s(T)\le B_i$ and $v_i(T)>v_i(X_i)$. The value of agent $i$'s current bundle is denoted by $u_i=v_i(X_i)$.

If $m=0$, the empty allocation is EF1. Henceforth $m\ge1$, and $n$ and $m$ always refer to the original instance. Agents may have zero utility, and goods that every agent values at zero may remain in charity.

\section{An algorithm for EF1 allocations}\label{sec:s3}

We begin with empty bundles and place all items in charity. Giving a single feasible item to an agent with an empty bundle preserves EF1 between agents. We use such assignments to give positive utility to agents whenever possible. The remaining updates replace an agent's bundle with a subset of charity. We construct this subset in one pass over the agents with positive utility, preserving each agent's fairness certificate as the candidate shrinks. Section~\ref{sec:s4} bounds the number of updates independently of the numerical magnitudes of the input.

\subsection{A sufficient condition for EF1}\label{sec:s3-1}

Fix an agent $i$ with current utility $u_i$ and a target set $D\subseteq M$. Write
\[
\OPT_i(D)=\max\{v_i(Y):Y\subseteq D,\ s(Y)\le B_i\}
\]
for the maximum value the agent can obtain from $D$. For nonempty $Y$, let $f_i(Y)=v_i(Y)-\max_{g\in Y}v_i(g)$, and set $f_i(\varnothing)=0$. Agent $i$ is EF1 towards $D$ precisely when $u_i\ge f_i(Y)$ for every feasible $Y\subseteq D$.

\begin{lemma}[Sufficient condition for EF1]\label{lem:ef1-sufficient}
If $u_i\ge(1-1/m)\OPT_i(D)$, then agent $i$ is EF1 towards $D$. In particular, let $0<\alpha\le1$ satisfy $\alpha^2\ge1-1/m$, and suppose a knapsack subroutine returns $K_i(D)$ satisfying
\begin{equation}\label{eq:knapsack}
K_i(D)\subseteq D,\qquad s(K_i(D))\le B_i,\qquad
v_i(K_i(D))\ge\alpha\OPT_i(D).
\end{equation}
Then $u_i\ge\alpha v_i(K_i(D))$ suffices for EF1 towards $D$. If $D$ itself is feasible for $i$, the single inequality $u_i\ge f_i(D)$ is necessary and sufficient.
\end{lemma}
\begin{proof}
For every nonempty feasible $Y\subseteq D$, its most valuable item is worth at least $v_i(Y)/|Y|$. Hence
\[
f_i(Y)\le\left(1-\frac1{|Y|}\right)v_i(Y)
\le\left(1-\frac1m\right)\OPT_i(D).
\]
This proves the first claim. By \eqref{eq:knapsack}, $\alpha v_i(K_i(D))\ge\alpha^2\OPT_i(D)$, giving the second. For the last claim, observe that $f_i$ is nondecreasing under inclusion. Adding an item to a nonempty set increases the residual by the smaller of the new item's value and the previous maximum, and adding the first item leaves the residual zero. Thus $f_i(Y)\le f_i(D)$ for every $Y\subseteq D$. When $D$ is feasible, the comparison with $Y=D$ also gives necessity.
\end{proof}

In the EF1 algorithm we take $\alpha=1-1/(2m)$ and write $\beta=\alpha^2$. Then $\beta=1-1/m+1/(4m^2)$ satisfies the assumption of Lemma~\ref{lem:ef1-sufficient}. We use a deterministic subroutine and evaluate the set it returns with the agent's original valuation. Items exceeding $B_i$ or having zero value to $i$ may be omitted from the call, and the subroutine returns the empty set if no items remain. We fix orders on agents and items to resolve ties.

The standard profit-scaling dynamic program~\cite[Section~8.2]{r27} returns a feasible set satisfying \eqref{eq:knapsack} in $O(m^4)$ arithmetic operations at this precision. Appendix~\ref{app:oracle} also describes how to recover a feasible set in the faster $\widetilde O(m+\eta^{-2})$ scheme of Chen et al.~\cite{r29} at relative error $\eta$. Throughout, $\widetilde O$ suppresses factors polylogarithmic in the combinatorial parameters and the inverse precision.

\subsection{The allocation algorithm}\label{sec:s3-2}

\paragraph{Activating agents with singletons.}
Call an agent \emph{active} once it has positive utility, and let $H$ denote the set of active agents. Before each search for a candidate, repeatedly assign an item $g\in C$ to an inactive agent $i$ whenever $s(g)\le B_i$ and $v_i(g)>0$. We call this a \emph{singleton activation}. Inactive agents have empty bundles, so an activation returns no items to charity. A singleton is automatically EF1 for every observer. All subsequent bundle replacements strictly increase the recipient's utility, so the set $H$ only grows. At most $m$ agents can ever be active, since each active agent holds a nonempty bundle and the bundles are disjoint.

When no further activation is possible, every inactive agent $i$ satisfies $\OPT_i(C)=0$. To see this, every item in a feasible subset of $C$ is itself feasible, because sizes are positive. Each such item must have value zero to $i$. Thus candidates contained in the current charity are automatically EF1 for all inactive agents, and only active agents need be processed.

\paragraph{Finding and shrinking a candidate.}
For each active agent $i$, test whether $\alpha v_i(K_i(C))>u_i$. The first successful test supplies an initial candidate $T=K_i(C)$ and a \emph{witness} $k=i$. If all tests fail, the stopping condition gives $u_i\ge\beta\OPT_i(C)$ for active agents; the same inequality holds for inactive agents because their charity optimum is zero.

Keep all utilities fixed while constructing a candidate. Process each active agent $i$ once. If $T$ does not fit $i$'s budget, call $K_i(T)$. A failed test $\alpha v_i(K_i(T))\le u_i$ certifies EF1 for every feasible subset of $T$, by Lemma~\ref{lem:ef1-sufficient}, so processing of $i$ ends. A successful test replaces $T$ with $K_i(T)$, records witness $i$, and makes $T$ feasible for $i$. For a feasible $T$, repeatedly remove a minimum-value item according to $v_i$ while $f_i(T)>u_i$, recording witness $i$ after each deletion. When this loop ends, the feasible-set case of the same lemma certifies all subsets of $T$. Either certificate remains valid as later agents shrink $T$, so there is no need to restart the scan.

Assign $T$ to its last witness and return that agent's old bundle to charity. The replacement is \emph{ordinary} if the last change to $T$ was its initialization or a successful knapsack call. It is a \emph{repair} if the last change was a minimum-value deletion.

\begin{algorithm}[H]
\caption{Computing an EF1 allocation by one-pass candidate construction}\label{alg:ef1}
\small
\begin{enumerate}[leftmargin=*,label=\arabic*.,itemsep=3pt,parsep=0pt,topsep=6pt]
\item Initialize $X_i=\varnothing$, $u_i=0$ for all $i$, $C=M$, and $H=\varnothing$. Set $\alpha=1-1/(2m)$.
\item While there exist $i\notin H$ and $g\in C$ with $s(g)\le B_i$ and $v_i(g)>0$, set $X_i=\{g\}$, $u_i=v_i(g)$, $H\leftarrow H\cup\{i\}$, and $C\leftarrow C\setminus\{g\}$.
\item Scan $i\in H$ in fixed order, computing $Z=K_i(C)$. At the first test with $\alpha v_i(Z)>u_i$, set $T=Z$, $k=i$, and mark the replacement ordinary. If all tests fail, return the allocation and charity.
\item Process each $i\in H$ once, keeping all utilities fixed:
\begin{enumerate}[leftmargin=*,label=(\alph*),itemsep=2pt]
\item If $s(T)>B_i$, compute $Z=K_i(T)$. If $\alpha v_i(Z)\le u_i$, continue with the next agent. Otherwise set $T=Z$, $k=i$, and mark the replacement ordinary.
\item Now $s(T)\le B_i$. While $f_i(T)>u_i$, choose $h\in\argmin_{g\in T}v_i(g)$, set $T\leftarrow T\setminus\{h\}$, $k=i$, and mark the replacement a repair.
\end{enumerate}
\item Let $X_k^{\mathrm{old}}=X_k$. Set $X_k=T$, $u_k=v_k(T)$, and $C\leftarrow(C\setminus T)\cup X_k^{\mathrm{old}}$. Keep all other bundles and return to Step 2.
\end{enumerate}
\noindent All choices follow fixed tie-breaking orders. Step 2 is completed before any candidate search. Section~\ref{sec:s4-3} implements it with item lists and a queue, without repeatedly scanning all inactive agents.
\end{algorithm}

\subsection{Correctness}\label{sec:s3-3}

\begin{lemma}[Candidate certificates and progress]\label{lemma:s3-1}
In Step 4 the candidate becomes strictly smaller at most $m-1$ times. The final candidate $T\subseteq C$ is nonempty and feasible for its last witness $k$. For every agent $i$ and every nonempty $Y\subseteq T$ with $s(Y)\le B_i$,
\begin{equation}\label{eq:candidate-certificate}
u_i\ge f_i(Y).
\end{equation}
Writing $u'_k=v_k(T)$, an ordinary replacement satisfies
\begin{equation}\label{eq:ordinary}
u'_k>u_k/\alpha.
\end{equation}
For a repair, let $P$ be the candidate immediately before its last deletion. Then $T$ is obtained from $P$ by deleting a minimum-value item for $k$, $1\le|T|\le m-1$, and
\begin{equation}\label{eq:repair}
\begin{aligned}
u'_k-u_k
&>\max_{g\in P}v_k(g)-\min_{g\in P}v_k(g)\\
&\ge\max_{g\in T}v_k(g)-\min_{g\in T}v_k(g)\ge0.
\end{aligned}
\end{equation}
\end{lemma}
\begin{proof}
The initial candidate has positive value for its witness and is feasible for that agent. A successful call in Step 4(a) produces a nonempty set feasible for its new witness. Since the previous candidate is infeasible for that agent, the returned subset is proper. Each deletion in Step 4(b) removes one item. The candidate remains nonempty because a singleton has residual zero and the loop stops before it can be deleted. Every shrink records a witness for whom the new candidate is feasible. Thus at most $m-1$ shrinks occur.

Consider the candidate $D$ at the end of processing an active agent $i$. If the knapsack test failed, then $u_i\ge\alpha v_i(K_i(D))$. Otherwise $D$ is feasible for $i$ and the deletion loop ended with $u_i\ge f_i(D)$. In either case, Lemma~\ref{lem:ef1-sufficient} gives $u_i\ge f_i(Y)$ for every feasible $Y\subseteq D$. The final candidate is contained in this $D$, so subsequent shrinks preserve the certificate. Inactive agents have $\OPT_i(C)=0$ after Step 2, and the candidate remains in that charity throughout construction; their certificates follow as well. This proves \eqref{eq:candidate-certificate} without assuming any monotonicity of the approximate oracle's output.

If the final replacement is ordinary, the last recorded witness satisfies $\alpha v_k(T)>u_k$, proving \eqref{eq:ordinary}. No subsequent shrink occurred, since each shrink would overwrite the witness and type. If it is a repair, its last deletion occurred at a set $P$ satisfying
\[
u_k<v_k(P)-\max_{g\in P}v_k(g).
\]
Deleting a minimum-value item leaves value $v_k(P)-\min_{g\in P}v_k(g)$, which proves \eqref{eq:repair}. Again, the last deletion produces the final $T$. Both types strictly increase the recipient's utility.
\end{proof}

\begin{theorem}[Correctness]\label{theorem:s3-2}
Algorithm~\ref{alg:ef1} terminates and returns a budget-feasible allocation that is EF1 between agents and EF1 towards charity. In addition, its final allocation satisfies
\begin{equation}\label{eq:charity-certificate}
v_i(X_i)\ge\beta\OPT_i(C)\qquad\text{for every }i.
\end{equation}
\end{theorem}
\begin{proof}
We maintain the partition into allocated bundles and charity, feasibility, inter-agent EF1, and the invariant that inactive agents have empty bundles. These statements hold initially. A singleton activation preserves feasibility and gives the recipient strictly positive utility. A singleton has residual zero for every observer, and the recipient's comparisons with all unchanged bundles only improve. Thus activation preserves inter-agent EF1 and the empty-bundle invariant for the remaining inactive agents.

For a replacement, Lemma~\ref{lemma:s3-1} shows that $T$ fits the recipient's budget and has greater value to it than its old bundle. Returning its old bundle to charity preserves the partition. Comparisons between unchanged agents are unaffected; the recipient's comparisons with unchanged bundles remain valid. Every other agent's comparisons with the new bundle follow from \eqref{eq:candidate-certificate}. Consequently, inter-agent EF1 is preserved. Positive utilities never decrease, so active agents remain active and inactive agents remain empty.

Each candidate construction is finite. Every actual assignment strictly increases the recipient's utility, and each agent has only finitely many subset values. Hence the algorithm terminates. This establishes termination independently of the polynomial bound proved next.

At termination, Step 2 has finished and all tests in Step 3 have failed. For active agents, these failed tests and \eqref{eq:knapsack} imply
$u_i\ge\alpha v_i(K_i(C))\ge\beta\OPT_i(C)$. For inactive agents, $\OPT_i(C)=0$. This proves \eqref{eq:charity-certificate}. Since $\beta\ge1-1/m$, Lemma~\ref{lem:ef1-sufficient} gives EF1 towards the charity for every agent.
\end{proof}

\section{Running time}\label{sec:s4}

Throughout, $n$ and $m$ are the numbers of agents and items in the original instance. We count repairs by identifying a utility threshold that each repair crosses. To count ordinary replacements, we bound the lengths of the intervals containing attainable positive utilities on a logarithmic scale. Both bounds are independent of the numerical magnitudes of the input.

Let $|I|$ denote the total binary encoding length of the input instance $I$. This includes the dimensions $n,m$ and the binary representations of all budgets, item sizes, and item values. Each rational entry is encoded by its numerator and denominator.

\begin{theorem}[Polynomial running time]\label{theorem:s4-1}
Algorithm~\ref{alg:ef1} is strongly polynomial. With the fast knapsack implementation in Appendix~\ref{app:oracle}, it uses
\[
\widetilde O(n^2m^4)\ \text{arithmetic operations}
\quad\text{and}\quad
\widetilde O\bigl(n^2m^4|I|^2\bigr)\ \text{bit operations}.
\]
Here $|I|$ is the binary input length, and the soft notation hides only factors polylogarithmic in $n$ and $m$. The standard profit-scaling implementation is also strongly polynomial and operates on integers of $O(|I|)$ bits.
\end{theorem}

To see why repairs need a separate count, consider four unit-size items of values $2H-1,H,H,H$, with one agent of budget $3$ and integer $H>4$. Suppose the first singleton assigned has value $2H-1$. All three remaining items fit together, and any two have value only $2H$. Since $\alpha=7/8$, the oracle must return all three, giving a candidate of value $3H$. Removing its most valuable item leaves $2H>2H-1$, so Step~4(b) deletes one item. The new bundle has value $2H$. The gain is only one, however large $H$ is. Yet the update crosses $2H$, the sum of two consecutive entries in the sorted list $H,H,H,2H-1$. The next lemma finds such a threshold for every repair.

\subsection{Bounding repairs by consecutive sums}\label{sec:s4-1}

\begin{lemma}[Consecutive-window certificate]\label{lemma:s4-2}
Let $a_1\le\cdots\le a_m$, and let $S\subseteq\{1,\ldots,m\}$ have cardinality $k\ge1$. Put $x=\sum_{j\in S}a_j$ and $R=\max_{j\in S}a_j-\min_{j\in S}a_j$. Some sum of $k$ consecutive entries satisfies
\[
x-R\le\sum_{j=t}^{t+k-1}a_j\le x.
\]
\end{lemma}
\begin{proof}
Write $S=\{i_1<\cdots<i_k\}$, $p=i_1$, and $q=i_k$. For $p\le t\le q-k+1$, set $c_t=\sum_{j=t}^{t+k-1}a_j$. The first window contains the $k$ smallest entries in $[p,q]$ and the last contains the $k$ largest, so $c_p\le x\le c_{q-k+1}$. Consecutive window sums satisfy
\[
0\le c_{t+1}-c_t=a_{t+k}-a_t\le a_q-a_p=R.
\]
Choose the last window in this range whose sum is at most $x$. Its sum is $x$, or its successor exceeds $x$ by a step of at most $R$. In either case its sum belongs to $[x-R,x]$.
\end{proof}

\begin{lemma}[Quadratic repair bound]\label{lemma:s4-3}
During any finite initial segment of the algorithm, each agent receives at most $m(m+1)/2-1$ repairs.
\end{lemma}
\begin{proof}
Fix agent $i$ and sort its original item values as $a_1\le\cdots\le a_m$, including zero values and repetitions. Define the fixed threshold set
\begin{equation}\label{eq:breakpoints}
D_i=\left\{\sum_{j=t}^{t+k-1}a_j:
1\le k\le m-1,\quad1\le t\le m-k+1\right\}.
\end{equation}
It contains at most $\sum_{k=1}^{m-1}(m-k+1)=m(m+1)/2-1$ distinct numbers. For a repair from $u$ to $u'=v_i(T)$, let $R$ be the range of item values on $T$. Lemma~\ref{lemma:s3-1} gives $1\le|T|\le m-1$ and $u'-u>R$. By Lemma~\ref{lemma:s4-2}, some $d\in D_i$ satisfies
\[
u<u'-R\le d\le u'.
\]
Every repair therefore crosses at least one threshold. Once the utility reaches $d$, it never falls below $d$ again. Hence a threshold in $(u,u']$ for one repair cannot be counted by a later repair. This proves the bound. When $m=1$, the threshold set is empty and repairs are impossible. The thresholds are used only for analysis and need not be constructed.
\end{proof}

\subsection{Bounding ordinary updates without valuation ratios}\label{sec:s4-2}

Count a singleton activation as an ordinary update for this subsection: it is the only possible update from zero utility, and it satisfies $u'>u/\alpha$. The other ordinary updates are exactly the ordinary replacements of Algorithm~\ref{alg:ef1}.

\begin{lemma}[Ordinary updates]\label{lemma:s4-4}
Fix an agent who values $q\ge1$ items positively. For any $0<\alpha<1$, in a nondecreasing sequence of subset utilities the number of updates satisfying $u'>u/\alpha$ is at most
\begin{equation}\label{eq:ordinary-count}
q+\frac{(q-1)\ln2}{-\ln\alpha}.
\end{equation}
In particular, for Algorithm~\ref{alg:ef1} this number is at most $m+2m(m-1)\ln2$. An agent with no positive item value has no updates.
\end{lemma}
\begin{proof}
Sort the positive values as $0<a_1\le\cdots\le a_q$, and write $A_j=\sum_{\ell=1}^j a_\ell$. Every positive subset sum lies in
\[
U=\bigcup_{j=1}^q[a_j,A_j]:
\]
choose $j$ as the largest index of a positive-valued item in the subset. On the logarithmic axis, this union has at most $q$ connected components and total length at most $(q-1)\ln2$. Indeed, the first interval is a point. Adding interval $j\ge2$ increases the covered logarithmic length by
\[
\ln\frac{A_j}{\max\{a_j,A_{j-1}\}}\le\ln2.
\]
If $a_j>A_{j-1}$, the interval starts a new component. Otherwise, because $a_j\ge a_{j-1}$, its overlap is already covered by the previous last interval, and only its extension beyond $A_{j-1}$ is new. Finally $A_j=A_{j-1}+a_j\le2\max\{A_{j-1},a_j\}$.

An update within a single positive component consumes the interval $(\ln u,\ln u']$, of length greater than $-\ln\alpha$. These intervals are disjoint because utilities never decrease. At most $q-1$ updates move between components, and at most one leaves zero. This proves \eqref{eq:ordinary-count}. For Algorithm~\ref{alg:ef1}, use $q\le m$ and $-\ln(1-1/(2m))\ge1/(2m)$.
\end{proof}

\subsection{Implementation and bit complexity}\label{sec:s4-3}

\paragraph{Singleton activations.}
In $O(nm)$ preprocessing, store for each item $g$ the list
\[
L_g=\{i:s(g)\le B_i,\ v_i(g)>0\}
\]
in fixed order, with a forward-only pointer. Initially put all items in a queue. Each time an old bundle is returned to charity, enqueue all its items. Implement Step 2 by emptying this queue. For a queued item still in charity, advance its pointer past active agents. Assign it to the first remaining inactive agent, if one exists; otherwise its list is exhausted and it stays in charity. Since agents only become active, an exhausted list can never gain an inactive agent. A charity item whose list has been exhausted therefore needs no further examination while it stays in charity. This proves that an empty queue implements the full stopping condition of Step 2.

Write $J$ for the total number of singleton activations and bundle replacements. Each list entry is skipped at most once, for $O(nm)$ total pointer work. Initially there are $m$ queue entries, and each replacement adds at most $m$, for $O((J+1)m)$ total queue operations besides the pointer advances. The ordered list of active agents has at most $m$ entries, and each of the $n$ agents enters it at most once. Insertions therefore take $O(nm)$ time in total. Steps~3 and~4 scan only this list.

\paragraph{Candidate construction.}
Step 3 makes at most one oracle call per agent, hence at most $n$ calls. Each agent in Step 4 makes at most one call: a failed test finishes its processing, and a successful test makes the candidate feasible for that agent. Thereafter its processing uses only exact residual tests and minimum-value deletions. Thus each replacement requires $O(n)$ oracle calls, including the initial search.

Once the candidate fits the agent's budget, compute its total value and the value of its most valuable item and build a min-heap of its items keyed by their value to the agent in $O(m)$ time. During minimum deletions the maximum value remains unchanged until a singleton is reached, when no further deletion is possible. Update the total value and size after each deletion. There are at most $m-1$ deletions over the entire construction, so the non-oracle work per replacement is $O(nm+m\log(m+1))$, including the manipulation of item sets. Updating the recipient's utility does not require recomputing unchanged agents' utilities. The final unsuccessful search makes at most $n$ calls.

\begin{proof}[Proof of Theorem~\ref{theorem:s4-1}]
Combining Lemmas~\ref{lemma:s4-3} and~\ref{lemma:s4-4}, each agent receives at most
\[
\frac{m(m+1)}2-1+m+2m(m-1)\ln2<2m^2
\]
updates, since $\ln 2<3/4$. Summing over the $n$ agents gives
\begin{equation}\label{eq:total-count}
J<2nm^2.
\end{equation}

Let $Q(m)$ be the cost of a call satisfying \eqref{eq:knapsack}. Adding preprocessing, activations, and all candidate constructions gives
\begin{equation}\label{eq:oracle-runtime}
O\!\left(nm+(J+1)\bigl(nQ(m)+nm+m\log(m+1)\bigr)\right).
\end{equation}
Since $J<2nm^2$, the algorithm makes $O(n^2m^2)$ knapsack calls in total. With the fast implementation, each call costs $Q(m)=\widetilde O(m^2)$ operations. Substitution in \eqref{eq:oracle-runtime} gives $\widetilde O(n^2m^4)$ operations, including the $O(nm)$ preprocessing cost. With standard profit scaling the same call count applies, with $Q(m)=O(m^4)$.

Clear denominators using one common scale for all sizes and budgets, and a separate scale for each agent's item values. These positive scalings preserve feasibility and all within-agent value comparisons. Each scale is a product of input denominators. Its binary length is at most the sum of the bit lengths of its factors, which is bounded by $|I|$. Scaling an entry replaces its own denominator in this product by its numerator, so each scaled entry also has $O(|I|)$ bits. Summing at most $m$ entries adds at most $\lceil\log_2 m\rceil$ bits. Thus bundle sizes and exact utilities have $O(|I|)$ bits, including after clearing the denominator in a comparison involving $\alpha=1-1/(2m)$.

For the standard dynamic program, the profit indices are bounded by $O(m^3)$ and the table entries are subset weights or a flag for an unattainable index. Thus the dynamic program also uses $O(|I|)$-bit integers. Clearing denominators costs $O(nm)$ arithmetic operations, using each common product once and dividing out the denominator of the entry being scaled. The call count and these operand bounds establish strong polynomiality for the standard implementation.

For the fast implementation, the weight and profit representations in Appendix~\ref{app:oracle} give an elementary bit cost per operation of $O(|I|^2)$ up to factors polylogarithmic in $m$. Summing over the fast operation count, including the $O(|I|)$ cost of reading the input, proves the stated bit bound.
\end{proof}

\section{Verifying EF1 allocations}\label{sec:s5}

We now study the complexity of checking whether a given allocation is EF1. All completeness statements in this section use polynomial-time many-one reductions.

\begin{definition}\label{definition:s5-1}
\EFValid{} asks whether a supplied feasible allocation $(X_1,\ldots,X_n,C)$ satisfies Definition~\ref{definition:s2-1}. Its complement, \EFInvalid{}, asks whether there are an agent $i$, a target bundle $D$ equal to another agent's bundle $X_j$ ($j\ne i$) or to the charity $C$, and a nonempty subset $X\subseteq D$ such that
\begin{equation}\label{eq:violation}
s(X)\le B_i,\qquad v_i(X)-\max_{g\in X}v_i(g)>v_i(X_i).
\end{equation}
\end{definition}

Partition and budget feasibility can be checked in polynomial time. The hardness results below hold even when these conditions are guaranteed for the supplied allocation.

\begin{theorem}[One agent and charity]\label{theorem:s5-2}
For one agent and a charity, \EFValid{} is coNP-complete.
\end{theorem}
\begin{proof}
A certificate for \EFInvalid{} consists of $(i,D,X)$. The subset, feasibility, and strict inequality in \eqref{eq:violation} can be checked in polynomial time, including for rational inputs. Thus \EFInvalid{} belongs to NP and \EFValid{} to coNP.

For hardness, reduce from the positive-integer \SSum{} problem. Given $a_1,\ldots,a_q$ and an integer $T\ge1$, decide whether some subset sums to $T$. This is a standard NP-complete problem~\cite{r28}. Construct one agent of budget $B=2T+1$. There are original items $g_1,\ldots,g_q$, an anchor item $h$, and an endowment item $d$, with
\[
\begin{array}{c|c|c}
\text{item}&\text{size}&\text{value}\\\hline
g_j&2a_j&a_j\\
h&1&1+\sum_{j=1}^q a_j\\
d&2T+1&T-1
\end{array}
\]
Supply the allocation
\[
X_1=\{d\},\qquad C=\{h,g_1,\ldots,g_q\}.
\]
The supplied allocation is feasible, and the construction has polynomial binary length.

Consider any nonempty feasible $X\subseteq C$. If $h\notin X$, then $s(X)=2v_1(X)\le2T+1$ implies $v_1(X)\le T$ by integrality. Since all original items have positive integer values,
\[
v_1(X)-\max_{g\in X}v_1(g)\le T-1=v_1(X_1).
\]
Thus such a subset cannot violate EF1.

If $h\in X$, write $X=\{h\}\cup\{g_j:j\in S\}$. The anchor is strictly more valuable than every original item, and
\[
s(X)\le2T+1\iff\sum_{j\in S}a_j\le T,\qquad
v_1(X)-\max_{g\in X}v_1(g)=\sum_{j\in S}a_j.
\]
Consequently, $X$ violates EF1 if and only if
\[
T-1<\sum_{j\in S}a_j\le T,
\]
which, by integrality, is equivalent to $\sum_{j\in S}a_j=T$. There are no comparisons with other agents. Hence
\[
(a_1,\ldots,a_q,T)\in\SSum\iff(X_1,C)\in\EFInvalid.
\]
Thus \EFInvalid{} is NP-complete and \EFValid{} is coNP-complete.

For any prescribed fixed $n>1$, first double all constructed sizes and the original budget. Add $n-1$ agents with budget 1, empty bundles, and the same valuation. Every item now has size at least 2, so the added agents have no nonempty feasible comparison subsets. The original agent's comparisons with their empty bundles are vacuous. The reduction is therefore preserved, with all budgets positive integers and all valuations identical.
\end{proof}

Assigning the comparison bundle to a second agent gives hardness with empty charity. The reduction below also proves that computing the maximum EF1 residual is NP-hard.

For an additive valuation $v$, the \emph{EF1 residual} of a nonempty set $X$ is the value remaining after removing an item of maximum value, namely $v(X)-\max_{g\in X}v(g)$. For a target bundle $D$ and budget $B$, the \emph{maximum EF1 residual} is
\[
R(D,B)=\max_{\substack{\varnothing\ne X\subseteq D\\s(X)\le B}}
\left(v(X)-\max_{g\in X}v(g)\right).
\]
We set $R(D,B)=0$ if there is no nonempty feasible subset of $D$. An agent with valuation $v$ and budget $B$ is EF1 towards $D$ precisely when the value of its own bundle is at least $R(D,B)$.

\begin{theorem}[Two agents and empty charity]\label{theorem:s5-3}
For two agents with identical valuations and an empty charity, \EFValid{} is coNP-complete. Computing the maximum EF1 residual $R(D,B)$ exactly is NP-hard.
\end{theorem}
\begin{proof}
Consider a 0--1 knapsack decision instance with items $g_j$ of positive integer sizes $b_j$, nonnegative integer profits $p_j$, budget $W\ge0$, and target profit $Q\ge1$. Let
\[
V^*=\max\left\{\sum_{j\in S}p_j:\sum_{j\in S}b_j\le W\right\}.
\]
Deciding whether $V^*\ge Q$ is NP-complete. Its hardness already follows from \SSum{} by setting $b_j=p_j=a_j$ and $W=Q=T$.

Give the original items sizes $2b_j$ and common values $p_j$. Add an anchor $h$ of size 1 and value $1+\sum_jp_j$, and an item $f$ of size 1 and value $Q-1$. Set
\[
B_1=2W+1,\quad B_2=2\sum_jb_j+1,\quad X_1=\{f\},\quad
X_2=\{h,g_1,\ldots,g_q\},\quad C=\varnothing.
\]
Both assigned bundles are feasible. Write $v$ for the common valuation. We claim
\begin{equation}\label{eq:residual}
R(X_2,B_1)=V^*.
\end{equation}
If a feasible $X$ contains $h$, its residual equals the profit of $X\setminus\{h\}$, whose original size is at most $W$. If $h\notin X$, its total constructed size is even. Thus $s(X)\le2W+1$ implies $s(X)\le2W$, and its residual is at most its profit, which is at most $V^*$. Conversely, adding $h$ to an optimal knapsack subset yields a feasible subset of $X_2$ with residual $V^*$. This proves \eqref{eq:residual}.

The only potentially violated comparison is agent 1 against $X_2$, since agent 2 compares with a singleton of zero residual and charity is empty. Therefore
\[
(X_1,X_2,C)\in\EFInvalid\iff V^*>Q-1\iff V^*\ge Q.
\]
The construction has polynomial binary length. Membership in coNP follows from the certificate argument in the proof of Theorem~\ref{theorem:s5-2}. Together with the reduction, this proves coNP-completeness of verification. Finally, identity \eqref{eq:residual} also reduces exact knapsack optimization to exact computation of $R$.
\end{proof}

Under polynomial-time Turing reductions, \EFValid{} is also NP-hard. Negating the answer from a validity oracle solves \EFInvalid{}.

\paragraph{Compatibility with polynomial-time construction.}
These results concern an arbitrary allocation supplied as part of the input. The algorithm in Theorems~\ref{theorem:s3-2} and~\ref{theorem:s4-1} constructs a particular EF1 allocation without solving either verification problem. Knapsack approximations provide sufficient candidate conditions, and the invariants certify each update. These conditions need not hold for an arbitrary EF1 allocation.

\section{Computing EFX allocations}\label{sec:s6}

EFX requires envy to disappear after the removal of \emph{any} item from a feasible comparison subset. Finding an EFX allocation is NP-hard even for one agent with charity. We give polynomial-time algorithms for valuations proportional to item size and for valuations with a fixed number of positive value types per agent. Both algorithms allow different budgets and include charity in the fairness guarantee.

\subsection{EFX and the stronger FEFx condition}\label{sec:s6-1}

\begin{definition}[EFX with charity]\label{definition:s6-1}
Consider a budget-feasible allocation. Agent $i$ is EFX towards agent $j$, where $j\ne i$, if, for every nonempty subset $X\subseteq X_j$ with $s(X)\le B_i$ and every $g\in X$, one has
\[
v_i(X_i)\ge v_i(X\setminus\{g\}).
\]
Agent $i$ is EFX towards the charity if the same inequality holds for every nonempty subset $X\subseteq C$ with $s(X)\le B_i$ and every $g\in X$. The allocation is EFX if every agent is EFX towards every other agent and every agent is EFX towards the charity. For additive valuations, each of these conditions is equivalent to
\begin{equation}\label{eq:efx}
v_i(X_i)\ge v_i(X)-\min_{g\in X}v_i(g)
\end{equation}
for every such subset $X$.
\end{definition}

For a target bundle $D$, which may be another agent's bundle or the charity, we also use the stronger condition
\begin{equation}\label{eq:fefx}
v_i(X_i)\ge v_i(Y)\quad\text{for every }Y\subsetneq D\text{ with }s(Y)\le B_i.
\end{equation}
Condition \eqref{eq:fefx}, imposed for every agent and target, is the FEFx notion of Barman et al.~\cite{r5}. It implies EFX as defined in Definition~\ref{definition:s6-1} because $X\setminus\{g\}$ is always a proper feasible subset of $D$. The converse need not hold. Consider one agent with budget 1, an empty assigned bundle, and two charity items with $(s,v)$ pairs $(1,1)$ and $(2,1)$. Every nonempty feasible subset is a singleton, so EFX holds. FEFx fails because the first item alone forms a proper feasible subset of the charity and has positive value.

\subsection{Hardness of finding an EFX allocation}\label{sec:efx-hardness}

\begin{theorem}[EFX search hardness]\label{thm:efx-search}
Finding an exact EFX allocation with charity is NP-hard under a polynomial-time Turing reduction that makes one oracle call. This holds even for one agent with a positive integer budget and positive integer item sizes and values encoded in binary.
\end{theorem}

\begin{proof}
Reduce from positive-integer \SSum{}~\cite{r28}. Given $a_1,\ldots,a_k$ and a target $t$, with $k,t\ge1$, construct a single-agent instance with a designated item $b_0$. Write $v$ for the agent's valuation and $X$ for its allocated bundle. We show that every EFX allocation satisfies
\begin{equation}\label{eq:efx-search-decode}
b_0\notin X
\quad\Longleftrightarrow\quad
\sum_{i\in I}a_i=t\text{ for some }I\subseteq\{1,\ldots,k\}.
\end{equation}

\smallskip
\noindent\textbf{Construction.}
Let $H=2^L$ be the smallest power of two strictly greater than $t+\sum_{i=1}^k a_i$, and set
\[
W=kH+t,\qquad B=2W+1,\qquad K=L+2.
\]
The agent has budget $B$. For each $i$, create an ordinary item of base weight $H+a_i$. Create $k$ more ordinary items of base weight $H$ and one special item $b_j$ for each $0\le j\le L$. For each $0\le j<L$, create three padding items of size $2^j$. Also create $\lfloor B/H\rfloor+1$ padding items of size $H$. With $D$ the total number of padding items, set $\Delta=D+1$ and assign sizes and values as follows.
\[
\begin{array}{lcc}
\hline
\text{Item} & \text{Size} & \text{Value}\\
\hline
\text{Ordinary item of base weight }w & 2w & \Delta K w\\
\text{Special item }b_j & B-2^j & \Delta(KW-j-1)\\
\text{Padding item of size }2^j & 2^j & 1\\
\hline
\end{array}
\]
All padding items together have value $D<\Delta$.

A subset of ordinary items has base weight exactly $W$ if and only if the \SSum{} instance is a YES instance. The base weight of $r$ ordinary items is $rH+\sum_{i\in I}a_i$, where $I$ indexes the chosen items of the form $H+a_i$. Both $t$ and $\sum_{i\in I}a_i$ are less than $H$, so equality with $kH+t$ forces $r=k$ and $\sum_{i\in I}a_i=t$. Conversely, any witness $I$ can be supplemented with $k-|I|$ of the weight-$H$ items.

\smallskip
\noindent\textbf{Decoding an arbitrary EFX allocation.}
Fix any EFX bundle $X$ and let $C=M\setminus X$. Since $B>2H$, every special item is larger than $B/2$. If $b_j\in X$, the remaining capacity is $2^j\le H$, while every ordinary item has size at least $2H$. Thus only padding items can accompany $b_j$, and
\begin{equation}\label{eq:efx-search-gap}
b_j\in X\quad\Longrightarrow\quad v(X)<v(b_j)+\Delta.
\end{equation}

\needspace{5\baselineskip}
\smallskip
\noindent\emph{First, $X$ cannot contain any $b_j$ with $j\ge1$.}
If $b_j\in X$, the remaining capacity $2^j$ accommodates at most two of the three pads of size $2^{j-1}$. Hence one such pad $p$ and the item $b_{j-1}$ both lie in $C$. The pair $\{b_{j-1},p\}$ has size exactly $B$. Applying EFX to this pair and deleting $p$ gives $v(X)\ge v(b_{j-1})=v(b_j)+\Delta$, contradicting \eqref{eq:efx-search-gap}.

\smallskip
\noindent\emph{Second, if $X$ contains no special item, the instance is YES.}
The pads of size $H$ have total size greater than $B$, so one of them, $p$, lies in $C$. Since $b_L$ also lies in $C$ and $s(\{b_L,p\})=B$, deleting $p$ from this pair gives the EFX inequality
\[
v(X)\ge v(b_L)=\Delta\bigl(K(W-1)+1\bigr).
\]
Let $w_X$ be the total base weight of ordinary items in $X$. Feasibility gives $w_X\le\lfloor B/2\rfloor=W$. If $w_X\le W-1$, then $v(X)\le\Delta K(W-1)+D<v(b_L)$. Thus $w_X=W$, and the ordinary items in $X$ encode a \SSum{} witness.

\smallskip
\noindent\emph{Third, on a YES instance $X$ cannot contain $b_0$.}
If $b_0\in X$, only one unit of capacity remains. All ordinary items and at least one unit pad $p$ lie in $C$. Choose a subset $S$ of ordinary items with total base weight $W$. The set $S\cup\{p\}$ has size $2W+1=B$. Deleting $p$ leaves value $v(S)=\Delta KW=v(b_0)+\Delta>v(X)$ by \eqref{eq:efx-search-gap}, violating EFX.

On a NO instance the first two claims force $b_0\in X$. On a YES instance the third forces $b_0\notin X$. This proves \eqref{eq:efx-search-decode}. In both cases a valid output exists because a maximum-value feasible bundle for one agent satisfies EFX. A \SSum{} algorithm calls the EFX oracle once and returns YES exactly when its output omits $b_0$.

Finally, $B/H=2k+(2t+1)/H<2k+2$ and $L=O(\log(1+t+\sum_i a_i))$. The construction lists $O(k+L)$ items with positive integer sizes and values of polynomial bit length. The construction and the test for membership of $b_0$ both take polynomial time.
\end{proof}

\subsection{Valuations proportional to item size}\label{sec:s6-2}

Suppose each agent $i$ has a positive coefficient $c_i$ such that
\begin{equation}\label{eq:proportional}
v_i(g)=c_i s(g)\qquad(i\in N,\ g\in M),\qquad c_i>0.
\end{equation}
Dividing agent $i$'s values by $c_i$ gives the common valuation $s$ and preserves all of its envy comparisons. For this equal-density case, Barman et al.~\cite[Theorem~2]{r2} show that their minimum-load greedy algorithm obtains EF1 with any consistent tie-breaking order on goods. We break ties by choosing the largest item that fits. Its size is then the smallest among all items assigned to that agent so far, which is precisely the deletion relevant to EFX.

\begin{algorithm}[H]
\caption{Minimum-load, largest-fitting allocation}\label{alg:proportional}
\small
Initialize $X_i=\varnothing$, loads $L_i=0$, remaining set $R=M$, and mark every agent active. While at least one agent remains active, perform the following steps.
\begin{enumerate}[leftmargin=*,label=\arabic*.]
\item Choose an active agent $i$ of minimum current load $L_i$.
\item If some $g\in R$ satisfies $L_i+s(g)\le B_i$, choose a maximum-size such item, assign it to $i$, set $L_i\leftarrow L_i+s(g)$, and remove it from $R$.
\item Otherwise, mark $i$ inactive permanently.
\end{enumerate}
Return $(X_1,\ldots,X_n,C)$, where $C=R$. All ties follow fixed orders.
\end{algorithm}

\begin{theorem}[Proportional valuations]\label{theorem:s6-2}
Under \eqref{eq:proportional}, Algorithm~\ref{alg:proportional} returns a feasible EFX allocation with charity using $O((n+m)^2)$ elementary operations on numbers of polynomial bit length.
\end{theorem}
\begin{proof}
Each iteration either assigns one item or deactivates one agent. There are at most $m+n$ iterations. Assignments preserve feasibility, and unassigned items are exactly the final charity.

First, the sizes in each agent's bundle are nonincreasing in order of assignment. Indeed, if an agent receives $a$ before $b$, then $b$ was still unassigned and also fit when $a$ was chosen, since the agent's load was no larger at that earlier time. The largest-fitting rule gives $s(a)\ge s(b)$.

Fix an observer $i$, and let $X_i$ denote its final bundle. Let $\tau_i$ be its deactivation time. No remaining item fits its residual budget $B_i-s(X_i)$ at $\tau_i$. Thus every item assigned after $\tau_i$, and every final charity item, has size strictly greater than $B_i-s(X_i)$.

Fix another agent $j$, and let $P$ be the prefix of $X_j$ assigned before $\tau_i$. If $P$ is nonempty, let $h$ be its last assigned item. At the time $h$ was assigned, agent $i$ was still active. Since $j$ was chosen with minimum active load,
\[
s(P\setminus\{h\})\le L_i\text{ at that time}\le s(X_i).
\]
The nonincreasing size order makes $h$ a minimum-size item in $P$, so
\begin{equation}\label{eq:prefix}
s(P)-\min_{g\in P}s(g)\le s(X_i).
\end{equation}
For nonempty sets, the total size remaining after deletion of a minimum-size item is monotone under inclusion. Indeed, adding an item of size at least the previous minimum increases this quantity by the new item's size. Adding a smaller item increases it by the previous minimum.

Now take any nonempty $X\subseteq X_j$ with $s(X)\le B_i$. If $X\subseteq P$, monotonicity and \eqref{eq:prefix} give
\[
s(X)-\min_{g\in X}s(g)\le s(P)-\min_{g\in P}s(g)\le s(X_i).
\]
Otherwise, the last-assigned item of $X$ was assigned after $\tau_i$. It is a minimum-size item of $X$, by the established assignment order, and its size exceeds $B_i-s(X_i)$. Hence
\[
s(X)-\min_{g\in X}s(g)\le B_i-\min_{g\in X}s(g)<s(X_i).
\]
For a nonempty feasible $X\subseteq C$, every item has size greater than $B_i-s(X_i)$, so the same inequality applies. If no such feasible subset exists, the comparison is vacuous.

Multiplying these size inequalities by $c_i$ yields \eqref{eq:efx}.

A direct implementation scans at most $n$ active loads and $m$ remaining items per iteration, giving $O((n+m)^2)$ operations. Loads and residual budgets are exact sums and differences of input sizes and budgets, with polynomial bit length. Hence the bit complexity is also polynomial.
\end{proof}

\subsection{A fixed number of positive value types}\label{sec:s6-3}

For agent $i$, let $a_{i,1},\ldots,a_{i,d_i}$ be the distinct positive values among $\{v_i(g):g\in M\}$, where $d_i$ is their number. Suppose $d_i\le d$ for every $i$, where $d\ge1$ is a fixed constant. The value levels, their magnitudes, and the items in each value class may all differ between agents. Values remain binary-encoded rationals. The bound concerns the number of levels, irrespective of their magnitudes.

We use the minimal-envied-set replacement framework of Barman et al.~\cite[Algorithms~2 and~3]{r5}. With a fixed number of positive value classes, we can solve each knapsack problem by enumerating how many items to take from each class. These same counts bound the number of distinct bundle values, and hence the number of times an agent's utility can increase.
\begin{lemma}[Exact knapsack with few value types]\label{lemma:s6-3}
An exact optimizer $K_i^*(D)$ for $\OPT_i(D)$ can be computed in
\[
O\bigl(m\log(m+1)+(d+1)(m+1)^d\bigr)
\]
elementary operations with polynomial bit complexity.
\end{lemma}
\begin{proof}
Partition the positive-valued items in $D$ into classes according to the levels $a_{i,1},\ldots,a_{i,d_i}$, allowing empty classes. Sort each class by increasing size and compute prefix size sums. For a count vector $(k_1,\ldots,k_{d_i})$, choosing the $k_t$ smallest items of class $t$ minimizes the total size among all subsets with those counts. Discard a count vector if it requests more items than its class contains. For each remaining vector, the subset of smallest total size is feasible exactly when the sum of the corresponding prefix sizes is at most $B_i$, and its value is
\[
\sum_{t=1}^{d_i}k_t a_{i,t}.
\]
Enumerate all vectors with coordinates in $\{0,\ldots,m\}$ and return a feasible subset of maximum value. Any optimal subset can have its zero-valued items removed without losing value or feasibility. The algorithm considers its count vector and chooses items of the same value and no greater total size. Thus the returned solution is exact.

Sorting and prefix sums take $O(m\log(m+1))$ operations, and there are at most $(m+1)^{d_i}\le(m+1)^d$ vectors, each tested in $O(d+1)$ time. Reconstructing the optimal subset costs $O(m)$. All sums and comparisons involve polynomial-bit rationals.
\end{proof}

\begin{algorithm}[H]
\caption{Exact minimal-candidate replacement}\label{alg:types}
\small
Start with $X_i=\varnothing$ and $C=M$. Repeat the following steps.
\begin{enumerate}[leftmargin=*,label=\arabic*.]
\item Set $u_i=v_i(X_i)$ for every $i$. Scan agents in the fixed order, computing $Z=K_i^*(C)$ for each agent $i$. At the first agent with $v_i(Z)>u_i$, set $T=Z$, $k=i$, and proceed to Step 2. If no agent passes the test, stop.
\item Keeping all utilities fixed, scan every pair $(i,h)$ with $h\in T$. If $Z=K_i^*(T\setminus\{h\})$ satisfies $v_i(Z)>u_i$, replace $T\leftarrow Z$, $k\leftarrow i$, and restart the scan. Continue until a complete scan makes no replacement.
\item Let $X_k^{\mathrm{old}}=X_k$. Assign the entire candidate $T$ to agent $k$ and return its old bundle to charity. Thus
\[
X_k\leftarrow T,\qquad C\leftarrow(C\setminus T)\cup X_k^{\mathrm{old}}.
\]
Keep all other bundles and repeat.
\end{enumerate}
\end{algorithm}

\begin{theorem}[A bounded number of distinct positive values]\label{theorem:s6-4}
If $d_i\le d$ for every agent and $d$ is fixed, Algorithm~\ref{alg:types} constructs an FEFx allocation, and therefore an EFX allocation, in time polynomial in the binary input length. The number of updates $J_{\mathrm{type}}$ satisfies
\begin{equation}\label{eq:type-count}
J_{\mathrm{type}}\le\sum_{i=1}^n\bigl((m+1)^{d_i}-1\bigr)\le n\bigl((m+1)^d-1\bigr).
\end{equation}
For fixed $d\ge1$, its elementary-operation count is
\[
O\bigl(n^2m^{2d+2}\log(m+1)\bigr).
\]
\end{theorem}
\begin{proof}
During candidate construction,
\[
T\subseteq C,\qquad s(T)\le B_k,\qquad v_k(T)>u_k
\]
hold initially and after every successful shrink. A shrink removes at least one item and leaves a nonempty candidate, so there are at most $m-1$ successful shrinks.

At the final failed scan, for every $i$ and $h\in T$,
\[
u_i\ge v_i(K_i^*(T\setminus\{h\}))=\OPT_i(T\setminus\{h\}).
\]
For any proper feasible $Y\subsetneq T$, choose $h\in T\setminus Y$ to obtain the exact certificate
\begin{equation}\label{eq:exact-certificate}
u_i\ge v_i(Y).
\end{equation}
Giving $T$ to its witness $k$ is feasible and strictly increases agent $k$'s utility. All comparisons between unchanged agents remain valid. The recipient's comparisons with unchanged bundles remain valid because its own utility increases. Every other agent's comparison with every proper feasible subset of the new bundle follows from \eqref{eq:exact-certificate}. Thus each replacement preserves FEFx between agents, and the initial empty bundles satisfy FEFx.

At termination, exact optimization gives
\[
v_i(X_i)\ge v_i(K_i^*(C))=\OPT_i(C).
\]
Thus no agent envies a feasible subset of charity, which implies FEFx towards charity. Zero-valued items do not affect this inequality.

We now bound the number of updates. For fixed $i$, every attainable bundle value is of the form
\[
\sum_{t=1}^{d_i}k_t a_{i,t},\qquad0\le k_t\le m.
\]
There are at most $(m+1)^{d_i}$ such values, including zero. Different count vectors can give the same value, which only decreases this number. Each update received by $i$ strictly increases its utility, and updates of other agents leave it unchanged. Starting at zero, agent $i$ can therefore receive at most $(m+1)^{d_i}-1$ updates. This proves \eqref{eq:type-count} for every finite prefix and hence also proves termination.

Each iteration uses at most $n$ initial oracle calls and at most $m$ scans of at most $nm$ pairs, hence $O(nm^2)$ exact knapsack calls. For fixed $d\ge1$, Lemma~\ref{lemma:s6-3} bounds each call by $O(m^d\log(m+1))$ operations, and \eqref{eq:type-count} bounds the number of iterations, including the stopping test, by $O(nm^d)$. Multiplying these three bounds gives the stated operation count. All arithmetic has polynomial bit complexity, yielding $n^2m^{O(d)}\operatorname{poly}(|I|)$ bit operations.
\end{proof}

\section{An application to social welfare under homogeneous valuations}\label{sec:s7}

We now assume that all agents share the same additive valuation $v$, while their budgets may still differ. The \emph{social welfare} of an allocation $A=(X_1,\ldots,X_n)$ is the total value assigned to agents,
\[
W(A)=\sum_{i=1}^n v(X_i).
\]
Let $W^\star$ be the maximum of this quantity over all budget-feasible allocations, without fairness constraints. This is the multiple-knapsack optimum for the given items and budgets. It differs from the sum of the optima of $n$ independently solved knapsacks, which might use the same item more than once.

Hu et al.~\cite[Theorems 3 and 4]{r6} prove that homogeneous instances admit EFX allocations with welfare at least $W^\star/2$, and give instances whose best EF1 welfare approaches this fraction as the number of agents grows. Their construction does not establish a polynomial running time in the binary input length. Moreover, the one-half argument assumes that each agent values its own bundle at least as much as every feasible subset of charity. This is stronger than EF1 towards charity. Our algorithm satisfies an approximate version of this property. We derive the resulting welfare bound, then adjust the approximation parameter to approach one half in polynomial time.

\subsection{From the charity condition to a welfare bound}\label{sec:s7-1}

\begin{lemma}[Welfare from a charity bound]\label{lemma:s7-1}
Suppose a budget-feasible allocation $A$ with charity $C$ satisfies, for some $0<\gamma\le1$,
\[
v(X_i)\ge\gamma\OPT_i(C)\quad\text{for every agent }i.
\]
Then
\[
W(A)\ge\frac{\gamma}{1+\gamma}W^\star.
\]
\end{lemma}
\begin{proof}
Fix an optimal allocation $(O_1,\ldots,O_n)$. Since $O_i\cap C$ is feasible for agent $i$, the assumed inequality gives
\[
v(O_i\cap C)\le\frac{v(X_i)}{\gamma}.
\]
The sets $O_i\setminus C$ are pairwise disjoint and contain only items allocated by $A$. Homogeneity and nonnegativity therefore give
\[
\begin{aligned}
W^\star
&=\sum_i v(O_i\cap C)+\sum_i v(O_i\setminus C)\\
&\le\frac{W(A)}{\gamma}+W(A).
\end{aligned}
\]
Rearranging proves the claim.
\end{proof}

At termination, Algorithm~\ref{alg:ef1} satisfies $v(X_i)\ge\beta\OPT_i(C)$ for every $i$. Thus its unmodified version guarantees
\begin{equation}\label{eq:welfare}
W(A)\ge\frac{\beta}{1+\beta}W^\star,\qquad\beta=\left(1-\frac{1}{2m}\right)^2.
\end{equation}
\subsection{Approaching one half in polynomial time}\label{sec:s7-2}

The approximation precision can be increased without changing the candidate construction or the repair rule. The repair bound is independent of this precision. Only the cost of knapsack calls and the number of ordinary updates change.

\begin{theorem}[EF1 with a near-half welfare guarantee]\label{thm:welfare-near-half}
For homogeneous nonnegative additive valuations and any rational $0<\varepsilon<1/2$, a budget-feasible EF1 allocation with charity and welfare at least $(1/2-\varepsilon)W^\star$ can be computed using
\[
\widetilde O\!\left(\frac{n^2m}{\delta^3}\right)
\quad\text{arithmetic operations},\qquad
\delta=\min\{1/(2m),\varepsilon\}.
\]
Let $|I_\varepsilon|$ be the total binary encoding length of the augmented input $(I,\varepsilon)$. The bit complexity is
\[
\widetilde O\!\left(
\frac{n^2m}{\delta^3}
|I_\varepsilon|^2
\right).
\]
With the standard profit-scaling oracle alone the bit complexity is also polynomial in $|I_\varepsilon|$ and $1/\varepsilon$.
\end{theorem}
\begin{proof}
Set $\delta=\min\{1/(2m),\varepsilon\}$, replace the initialization of $\alpha$ in Algorithm~\ref{alg:ef1} by $\alpha=1-\delta$, and set $\beta=\alpha^2$. Both implementations in Appendix~\ref{app:oracle} return a feasible set worth at least $(1-\delta)\OPT_i(D)$.

Since $\delta\le1/(2m)$, we have $\beta\ge(1-1/(2m))^2\ge1-1/m$. Thus Lemma~\ref{lem:ef1-sufficient} applies at this precision, and all candidate certificates remain valid. Singleton activations are independent of the precision. The proof of Theorem~\ref{theorem:s3-2} therefore gives exact EF1 and the stopping condition $v(X_i)\ge\beta\OPT_i(C)$ for all agents. Lemma~\ref{lemma:s7-1} gives welfare at least $\beta W^\star/(1+\beta)$, and
\[
\frac12-\frac{\beta}{1+\beta}
=\frac{1-\beta}{2(1+\beta)}
=\frac{2\delta-\delta^2}{2(1+\beta)}\le\delta\le\varepsilon.
\]

The repair bound in Lemma~\ref{lemma:s4-3} does not depend on $\alpha$. For ordinary updates, Lemma~\ref{lemma:s4-4} and $-\ln(1-\delta)\ge\delta$ give at most $m+(m-1)\ln2/\delta$ per agent, counting its activation. Hence
\[
J=O\!\left(n(m^2+m/\delta)\right).
\]
Because $\delta\le 1/(2m)$, we have $J=O(nm/\delta)$. With the standard oracle, \eqref{eq:oracle-runtime} and $Q(m)=O(m^3/\delta)$ give $O(n^2m^4/\delta^2)$ arithmetic operations. All table indices and exact comparisons have polynomial bit length in $|I_\varepsilon|$, proving the polynomial-time welfare guarantee independently.

With the fast oracle, $Q(m)=\widetilde O(m+\delta^{-2})$. Substitution in \eqref{eq:oracle-runtime} gives $\widetilde O(n^2m/\delta^3)$ arithmetic operations. Clearing denominators as in Section~\ref{sec:s4-3} and using the representations in Appendix~\ref{app:oracle} bounds the elementary bit cost per operation by $O(|I_\varepsilon|^2)$ up to factors polylogarithmic in $m$ and $1/\delta$. This proves the displayed bit bound.
\end{proof}

Hu et al.~\cite[Theorem 4]{r6} construct homogeneous instances with identical budgets in which the ratio of the best EF1 welfare to the maximum feasible welfare approaches $1/2$ as the number of agents grows. Thus no constant welfare guarantee greater than $1/2$ can hold uniformly over all numbers of agents, making the limiting fraction in Theorem~\ref{thm:welfare-near-half} optimal.

\section*{Declaration of Use of Generative AI}

The development of this work involved extensive and iterative discussions with OpenAI's GPT6 Astra. The authors formulated the research questions, contributed mathematical and algorithmic ideas, identified relevant techniques and connections to previous research, and guided the overall direction of the investigation. ChatGPT assisted in organizing the presentation, and preparing portions of the manuscript in \LaTeX.

\appendix
\section{Constructive approximate knapsack}\label{app:oracle}

Consider one knapsack call on a set $D$ of at most $m$ items. Write $w(g)$ for the size of item $g$ and $w(S)$ for the total size of a set $S$. Discard items that individually exceed the capacity or have zero profit. If none remain, return the empty set. Write $p_g$ for the positive profit of a remaining item, $P=\max_g p_g$, and $V$ for the optimum of the call. In particular, $P\le V$. Let $0<\eta<1$ be the desired relative error. The weights and capacity are cleared of denominators as in Section~\ref{sec:s4-3}.

\paragraph{Profit scaling and the standard implementation.}
Take $K=\eta P/m$ and $q_g=\lfloor p_g/K\rfloor$. For every feasible $S$,
\[
 K\sum_{g\in S}q_g\le p(S)\le K\sum_{g\in S}q_g+mK .
\]
For each prefix of the item list, let $D_j$ be the minimum weight attaining rounded profit exactly $j$, with $+\infty$ for an unattainable index. Initially $D_0=0$ and $D_j=+\infty$ for $j>0$. On processing $g$ set
\[
 D'_j=\min\{D_j,\,w(g)+D_{j-q_g}\},
\]
where $D_k=+\infty$ for $k<0$. Both entries on the right come from the preceding row, so this recurrence also handles $q_g=0$. At each finite entry, record whether the recurrence took the item. Trace back from the largest final index $j$ whose stored weight is at most the capacity. For the resulting feasible set $Z$ and an optimal set $S^\star$, we have $p(Z)\ge Kq(Z)\ge Kq(S^\star)\ge V-mK\ge(1-\eta)V$, where $q(S)=\sum_{g\in S}q_g$. There are at most $1+\lfloor m^2/\eta\rfloor$ indices per row and at most $m$ item rows. This gives $O(m^3/\eta)$ operations, or $O(m^4)$ at $\eta=1/(2m)$.

\paragraph{Fast implementation and recovery of a set.}
First apply the same scaling with $K=\eta P/(4m)$. The rounded profits lie in $[0,4m/\eta]$. Discard goods with zero rounded profit and apply the deterministic fully polynomial-time approximation scheme (FPTAS) of Chen et al.~\cite{r29} to the others at accuracy $\eta/2$. Its approximation factor is $1+\eta/2$. Any feasible set $Z$ attaining the reported lower bound therefore satisfies
\[
 p(Z)\ \ge\ Kq(Z)\ \ge\ \frac{V-mK}{1+\eta/2}
 \ \ge\ \frac{1-\eta/4}{1+\eta/2}V\ \ge\ (1-\eta)V .
\]
Initial scaling takes $O(m)$ operations. To implement the $\widetilde O(m+\eta^{-2})$ scheme constructively, label the original items $0,\ldots,m-1$. Alongside a finite table entry store an integer
\[
 E(S)=\sum_{g\in S}2^g
\]
to identify the set $S$ that realizes the entry. Compute the $m$ singleton codes once by doubling. These codes are auxiliary data used only to recover the items.

For a packing entry at capacity $x$ with reported profit $y$, its set $S$ has weight at most $x$ and profit at least $y$ in the current units. Rounding the reported profit down or imposing an upper limit on it preserves this guarantee. When taking a pointwise maximum of profit functions or choosing among dynamic programming entries, copy the code from an entry that gives the selected value. In a convolution, record the split attaining the selected rounded entry as in Chan~\cite[Section~5]{r30}. The item groups of the two children are disjoint, so adding their codes gives the code of their union. Enumerating the splits in a general rounded two-function merge takes $\widetilde O(L+\xi^{-2})$ operations for inputs of total step complexity $L$ and local precision $\xi$~\cite[Fact~1(i) and Lemmas~1--2]{r30}. For a class of items with the same profit, the matrix search returns the number of items to take from that class. Store codes alongside the prefix weight sums, in increasing weight order for packing and decreasing order for covering.

For a prefix $G$ of items ordered by decreasing profit-to-weight ratio, the subproblem asks which items to remove. Their total weight must meet a given demand, and their total profit should be as small as possible~\cite[Section~6.2]{r29}. Round a covering loss up. An entry for demand $d$ with reported loss $\ell$ then represents a set $U\subseteq G$ of weight at least $d$ and profit at most $\ell$. For packing capacity $x$, use $d=w(G)-x$ and take $U=\varnothing$ if $d\le0$. Writing $p_G$ for profit in the units of this prefix, we obtain
\[
 w(G\setminus U)\le x,\qquad p_G(G\setminus U)\ge p_G(G)-\ell .
\]
The complement has code $E(G)-E(U)$. For a loss cutoff $t$ and relative upper error $\xi$, keep entries through $(1+\xi)t$, so a rounded loss for a true loss of $t$ is retained. If a constant was added to all profits in a table before merging, subtract it when interpreting the result. These are the packing and covering conventions in~\cite[Appendices~A and~D]{r31}.

A grouped item carries the sum of its constituents' codes. Prefix codes for the nested high-efficiency prefixes are computed cumulatively, together with their weights. Each choice, merge, or complementation thus adds only a constant number of arithmetic operations per entry already computed. Reading the $m$ binary digits of the final code by successive divisions by two recovers $Z$ in $O(m)$ operations. This preserves the $\widetilde O(m+\eta^{-2})$ arithmetic bound.

\paragraph{Bit cost.}
Let $b$ be the largest bit length among the cleared weights, capacity, profits, and the numerator and denominator of $\eta$, plus $\lceil\log_2(m+2)\rceil$ bits for sums. Computing the $q_g$ uses products and divisions of a constant number of $O(b)$-bit integers. After scaling, the total integer profit is at most $4m^2/\eta$. Weight coordinates remain exact. A finite penalty used in a matrix search is bounded by a weight sum times a polynomial in the index range~\cite[Appendix~A]{r31}. Each set code is less than $2^m$.

Put $h=\lceil\log_2(m+2+1/\eta)\rceil$. In the normalization of~\cite[Section~4]{r29}, use the profit of a feasible constant-factor solution and round the internal inverse precision up to a power of four. The feasible solution has scaled profit at most $4m^2/\eta$. Even after the polylogarithmic precision adjustment, the chosen power of four has $O(h)$ bits and an integer square root. The profit-band rescalings therefore use $O(h)$-bit rationals. In a dyadic profit band $[v/2,v]$, use an additive mesh of order $\xi v$ for relative precision $\xi$, retaining a breakpoint's witness when its profit is rounded. The rounding error is $O(\xi)$ relative within that band. Choosing the constant in the mesh within the merge's error budget gives the prescribed precision. For a structured merge, use the grid aligned to the common item profit and recompress at the output of the merge~\cite[Appendix~A.1]{r31}. The normalized bands span a polynomial range in $m$ and $1/\eta$. Their exponents have magnitude $O(h)$, and the local inverse precisions and table sizes are polynomial in $m$ and $1/\eta$. An index, its grid spacing, and a band origin specify a stored profit. Recompression resets the denominator at each merge. Efficiency comparisons use cross-products of a weight and a profit.

The numerical operands thus use $O(b+\operatorname{poly}(h))$ bits, and the set codes use at most $m$ bits each. The explicit input lists all $m$ items. At $\eta=1/(2m)$ we have $m,b=O(|I|)$ and $h=O(\log(m+2))$. Schoolbook arithmetic therefore takes $O(|I|^2)$ bit operations per arithmetic operation up to factors polylogarithmic in $m$. At $\eta=\delta$ in Theorem~\ref{thm:welfare-near-half}, the same argument gives $O(|I_\varepsilon|^2)$ up to factors polylogarithmic in $m$ and $1/\delta$.


\begin{thebibliography}{31}
\bibitem{r1} Edith Elkind, Ayumi Igarashi, and Nicholas Teh. Fair division of chores with budget constraints. In Proceedings of the International Symposium on Algorithmic Game Theory, 2024. Full version: arXiv:2410.23979.

\bibitem{r2} Siddharth Barman, Arindam Khan, Sudarshan Shyam, and K. V. N. Sreenivas. Finding fair allocations under budget constraints. Proceedings of the AAAI Conference on Artificial Intelligence, 37(5):5481--5489, 2023. \href{https://doi.org/10.1609/aaai.v37i5.25681}{\nolinkurl{doi:10.1609/aaai.v37i5.25681}}.

\bibitem{r3} Eric Budish. The combinatorial assignment problem: Approximate competitive equilibrium from equal incomes. Journal of Political Economy, 119(6):1061--1103, 2011. \href{https://doi.org/10.1086/664613}{\nolinkurl{doi:10.1086/664613}}.

\bibitem{r4} Richard J. Lipton, Evangelos Markakis, Elchanan Mossel, and Amin Saberi. On approximately fair allocations of indivisible goods. In Proceedings of the ACM Conference on Electronic Commerce, pages 125--131, 2004. \href{https://doi.org/10.1145/988772.988792}{\nolinkurl{doi:10.1145/988772.988792}}.

\bibitem{r5} Siddharth Barman, Arindam Khan, Sudarshan Shyam, and K. V. N. Sreenivas. Guaranteeing envy-freeness under generalized assignment constraints. In Proceedings of the ACM Conference on Economics and Computation, pages 242--269, 2023. \href{https://doi.org/10.1145/3580507.3597698}{\nolinkurl{doi:10.1145/3580507.3597698}}. Full version: arXiv:2305.01339.

\bibitem{r6} Tingwei Hu, Lili Mei, Zhen Wang, and Guochuan Zhang. Nearly tight bounds on the price of fairness for indivisible items under budget constraints. Theoretical Computer Science, 1080:116036, 2026. \href{https://doi.org/10.1016/j.tcs.2026.116036}{\nolinkurl{doi:10.1016/j.tcs.2026.116036}}.

\bibitem{r7} Ioannis Caragiannis, David Kurokawa, Hervé Moulin, Ariel D. Procaccia, Nisarg Shah, and Junxing Wang. The unreasonable fairness of maximum Nash welfare. ACM Transactions on Economics and Computation, 7(3):12:1--12:32, 2019. \href{https://doi.org/10.1145/3355902}{\nolinkurl{doi:10.1145/3355902}}.

\bibitem{r8} Benjamin Plaut and Tim Roughgarden. Almost envy-freeness with general valuations. SIAM Journal on Discrete Mathematics, 34(2):1039--1068, 2020. \href{https://doi.org/10.1137/19M124397X}{\nolinkurl{doi:10.1137/19M124397X}}.

\bibitem{r9} Bhaskar Ray Chaudhury, Jugal Garg, and Kurt Mehlhorn. EFX exists for three agents. In Proceedings of the ACM Conference on Economics and Computation, pages 1--19, 2020. Full version: arXiv:2002.05119.

\bibitem{r10} Bhaskar Ray Chaudhury, Telikepalli Kavitha, Kurt Mehlhorn, and Alkmini Sgouritsa. A little charity guarantees almost envy-freeness. SIAM Journal on Computing, 50(4):1336--1358, 2021. \href{https://doi.org/10.1137/20M1359134}{\nolinkurl{doi:10.1137/20M1359134}}.

\bibitem{r11} Ben Berger, Avi Cohen, Michal Feldman, and Amos Fiat. Almost full EFX exists for four agents. Proceedings of the AAAI Conference on Artificial Intelligence, 36(5):4826--4833, 2022. \href{https://doi.org/10.1609/aaai.v36i5.20410}{\nolinkurl{doi:10.1609/aaai.v36i5.20410}}.

\bibitem{r12} Ioannis Caragiannis, Nick Gravin, and Xin Huang. Envy-freeness up to any item with high Nash welfare: The virtue of donating items. In Proceedings of the ACM Conference on Economics and Computation, pages 527--545, 2019. Full version: arXiv:1902.04319.

\bibitem{r13} Michal Feldman, Simon Mauras, and Tomasz Ponitka. On optimal tradeoffs between EFX and Nash welfare. Proceedings of the AAAI Conference on Artificial Intelligence, 38(9):9688--9695, 2024. \href{https://doi.org/10.1609/aaai.v38i9.28826}{\nolinkurl{doi:10.1609/aaai.v38i9.28826}}.

\bibitem{r14} Xiaowei Wu, Bo Li, and Jiarui Gan. Budget-feasible maximum Nash social welfare is almost envy-free. In Proceedings of the International Joint Conference on Artificial Intelligence, pages 465--471, 2021. \href{https://doi.org/10.24963/ijcai.2021/65}{\nolinkurl{doi:10.24963/ijcai.2021/65}}.

\bibitem{r15} Jiarui Gan, Bo Li, and Xiaowei Wu. Approximation algorithm for computing budget-feasible EF1 allocations. In Proceedings of the International Conference on Autonomous Agents and Multiagent Systems, pages 170--178, 2023.

\bibitem{r16} Xiaowei Wu, Bo Li, and Jiarui Gan. Approximate envy-freeness in indivisible resource allocation with budget constraints. Information and Computation, 303:105264, 2025. \href{https://doi.org/10.1016/j.ic.2024.105264}{\nolinkurl{doi:10.1016/j.ic.2024.105264}}.

\bibitem{r17} Saar Cohen, Nicholas Teh, Paul W. Goldberg, and Michael J. Wooldridge. Online fair division with budget constraints. arXiv:2607.23310, 2026.

\bibitem{r18} Warut Suksompong. Constraints in fair division. ACM SIGecom Exchanges, 19(2):46--61, 2021.

\bibitem{r19} Arpita Biswas and Siddharth Barman. Fair division under cardinality constraints. In Proceedings of the International Joint Conference on Artificial Intelligence, pages 91--97, 2018. \href{https://doi.org/10.24963/ijcai.2018/13}{\nolinkurl{doi:10.24963/ijcai.2018/13}}.

\bibitem{r20} Dimitris Bertsimas, Vivek F. Farias, and Nikolaos Trichakis. The price of fairness. Operations Research, 59(1):17--31, 2011. \href{https://doi.org/10.1287/opre.1100.0865}{\nolinkurl{doi:10.1287/opre.1100.0865}}.

\bibitem{r21} Ioannis Caragiannis, Christos Kaklamanis, Panagiotis Kanellopoulos, and Maria Kyropoulou. The efficiency of fair division. Theory of Computing Systems, 50(4):589--610, 2012. \href{https://doi.org/10.1007/s00224-011-9359-y}{\nolinkurl{doi:10.1007/s00224-011-9359-y}}.

\bibitem{r22} Xiaohui Bei, Xinhang Lu, Pasin Manurangsi, and Warut Suksompong. The price of fairness for indivisible goods. Theory of Computing Systems, 65(7):1069--1093, 2021. Full version: arXiv:1905.04910.

\bibitem{r23} Siddharth Barman, Umang Bhaskar, and Nisarg Shah. Optimal bounds on the price of fairness for indivisible goods. In Proceedings of the International Conference on Web and Internet Economics, pages 356--369, 2020. Full version: arXiv:2007.06242.

\bibitem{r24} Zihao Li, Shengxin Liu, Xinhang Lu, Biaoshuai Tao, and Yichen Tao. A complete landscape for the price of envy-freeness. In Proceedings of the International Conference on Autonomous Agents and Multiagent Systems, pages 1183--1191, 2024. Full version: arXiv:2401.01516.

\bibitem{r25} Haris Aziz, Xin Huang, Nicholas Mattei, and Erel Segal-Halevi. Computing welfare-maximizing fair allocations of indivisible goods. European Journal of Operational Research, 307(2):773--784, 2023. Full version: arXiv:2012.03979.

\bibitem{r26} Xiaolin Bu, Zihao Li, Shengxin Liu, Jiaxin Song, and Biaoshuai Tao. Approximability landscape of welfare maximization within fair allocations. In Proceedings of the ACM Conference on Economics and Computation, pages 412--440, 2025. Full version: arXiv:2205.14296.

\bibitem{r27} Vijay V. Vazirani. Approximation algorithms. Springer, 2001. \href{https://doi.org/10.1007/978-3-662-04565-7}{\nolinkurl{doi:10.1007/978-3-662-04565-7}}.

\bibitem{r28} Richard M. Karp. Reducibility among combinatorial problems. In Complexity of Computer Computations, pages 85--103. Plenum Press, 1972. \href{https://doi.org/10.1007/978-1-4684-2001-2_9}{\nolinkurl{doi:10.1007/978-1-4684-2001-2_9}}.
\bibitem{r29} Lin Chen, Jiayi Lian, Yuchen Mao, and Guochuan Zhang. A nearly quadratic-time FPTAS for knapsack. SIAM Journal on Computing, published online August 20, 2025. \href{https://doi.org/10.1137/24M1686772}{\nolinkurl{doi:10.1137/24M1686772}}. Full version: \href{https://arxiv.org/abs/2308.07821}{arXiv:2308.07821}.

\bibitem{r30} Timothy M. Chan. Approximation schemes for 0-1 knapsack. In 1st Symposium on Simplicity in Algorithms (SOSA 2018), volume 61 of Open Access Series in Informatics, pages 5:1--5:12, 2018. \href{https://doi.org/10.4230/OASIcs.SOSA.2018.5}{\nolinkurl{doi:10.4230/OASIcs.SOSA.2018.5}}.

\bibitem{r31} Lin Chen, Jiayi Lian, Yuchen Mao, and Guochuan Zhang. A note on truly subquadratic time FPTAS for knapsack. \href{https://arxiv.org/abs/2611.22809}{arXiv:2611.22809}, 2026.
\end{thebibliography}
\end{document}